\documentclass{ifacconf}
\usepackage{packages_and_commands}
\begin{document}
\begin{frontmatter}

\title{A Simulation Preorder for Koopman-like Lifted Control Systems\thanksref{footnoteinfo}}
\thanks[footnoteinfo]{This work is funded by the ONR grant N00014-21-1-2431 (CLEVR-AI).}

\author[First]{Antoine Aspeel} 
\author[First]{Necmiye Ozay}
\address[First]{Electrical Engineering and Computer Science Department, Univ. of Michigan, Ann Arbor, MI (e-mail: \{antoinas,necmiye\}@umich.edu).}

\begin{abstract}                
This paper introduces a simulation preorder among lifted systems, a generalization of finite-dimensional Koopman approximations (also known as approximate immersions) to systems with inputs. It is proved that this simulation relation implies the containment of both the open- and closed-loop behaviors. Optimization-based sufficient conditions are derived to verify the simulation relation in two special cases: i) a nonlinear (unlifted) system and an affine lifted system and, ii) two affine lifted systems. Numerical examples demonstrate the approach.
\end{abstract}

\begin{keyword}
Nonlinear control, Simulation relation, Safety control, Immersion. 
\end{keyword}

\end{frontmatter}

\section{Introduction}

Nonlinear systems are ubiquitous in nature and engineering, exhibiting complex dynamics that often defy linear modeling approaches. The Koopman operator framework \citep{koopman1931hamiltonian}, offers a promising alternative for understanding and analyzing nonlinear systems. This framework transforms a finite-dimensional nonlinear system into an infinite-dimensional linear system by viewing it through the lens of an infinite set of observable functions. While the Koopman operator provides a powerful theoretical foundation, its practical application is hindered by its infinite-dimensional nature. Indeed, nonlinear systems generally do not admit an exact finite-dimensional linear representation \citep{levine1986nonlinear,liu2023non}.


To address these limitations, finite-dimensional Koopman approximations --- also known as approximate immersions \citep{wang2023computation} --- have emerged as a practical tool for modeling and analyzing nonlinear systems \citep{williams2015data}. These approaches lift the state into a higher --- but finite --- dimensional space through a lifting function. A linear dynamics in the lifted space is constructed to approximate the nonlinear dynamics. In \citep{sankaranarayanan2016change}, an autonomous system is abstracted into a linear (or algebraic) lifted system. This is used to infer invariants of the nonlinear system. In \citep{wang2023computation}, an autonomous nonlinear system is lifted into a linear one to compute invariant sets. When such an immersion does not exist, a method to compute approximate immersions is proposed. Alternatively, extended dynamic mode decomposition (EDMD) has been proposed as a data-driven algorithm that computes a Koopman approximation using a dictionary of functions, which defines the lifting function \citep{williams2015data}. However, there is no guarantee that EDMD finds an accurate approximation when the subspace spanned by the dictionary is not invariant under the Koopman operator. To address this limitation, the Tunable Symmetric Subspace Decomposition algorithm is introduced in \citep{haseli2023generalizing} to trade off between invariance and expressiveness by pruning the dictionary. While EDMD was originally developed for autonomous systems, generalizations to systems with inputs have been developed \citep{proctor2018generalizing}. For example, in \citep{korda2018linear} model predictive control on a linear lifted system is used to control a nonlinear system. However, one major challenge with EDMD is the selection of the lifting function, which can significantly impact the accuracy and applicability of the approximation. In \cite[Sec.~6.4]{colbrook2023multiverse}, this is considered ``one of the most significant open problems" in the finite-dimensional Koopman framework.

In this work, we develop a theoretical framework to tackle this challenge. Inspired by the success of simulation relations to build a simpler (e.g., discrete) representation of a nonlinear system \citep{girard2007approximation, tabuada2009verification, zamani2011symbolic, liu2016finite}, we propose a new notion of simulation between lifted control systems.

\emph{Contribution:} We introduce the concept of a lifted system as a natural generalization of finite-dimensional Koopman models to lifted nonlinear and set-valued dynamical systems with inputs. Then, a notion of simulation between lifted systems is defined. It is proved that when one lifted system simulates another, the open- (resp. closed-) loop behavior of the former is contained within the open- (resp. closed-) loop behavior of the latter. It follows that if a simulating lifted system satisfies some given specifications for a given controller, then the same controller can be used to satisfy the same specifications for the simulated system. In addition, this new notion of simulation relation is shown to unify and generalize Koopman over-approximations~\citep{balim2023koopman}, approximate immersions~\citep{wang2023computation}, and hybridizations~\citep{girard2011synthesis}. Finally, we derive optimization-based sufficient conditions to (i) find an affine lifted system that simulates a nonlinear polynomial system, and (ii) verify if one affine lifted system simulates another. These results enable us to compare different lifting functions and alternative lifted systems in terms of their usefulness in control design. Our framework is illustrated on numerical examples.

\emph{Notation}: For two sets $\X$ and $\Y$, $f:\X\rightarrow\Y$ denotes a function from $\X$ to $\Y$, and $f:\X\rarrows\Y$ denotes a set-valued map from $\X$ to $\Y$, i.e., it is a function from $\X$ to $2^\Y$. For a subset $X\subseteq\X$, the image of $X$ under a function $f:\X\rightarrow\Y$ is the set $f(X)\coloneqq\{f(x)\in\Y\mid x\in X\}$. For a subset $Y\subseteq\Y$, the pre-image of $Y$ under $f$ is the set $f^{-1}[Y]\coloneqq\{x\in\X\mid f(x)\in Y\}$. Intervals in $\R$ are denoted $[a,b]$ and discrete intervals are denoted $[a;b]\coloneqq[a,b]\cap\mathbb{Z}$, where $\mathbb{Z}$ is the set of integers. For a set $\A$, $\A^*\coloneqq \bigcup_{T\in\mathbb{Z}_+}\A^{[0;T[}$ is the set of finite signals with values in $\A$, while $\A^\infty\coloneqq \A^*\cup\A^{[0;\infty[}$ is the set of finite and infinite signals with values in $\A$. The Minkowski sum of two sets $\X$ and $\Y$ is denoted $\X\oplus\Y$, and the $n$-th Minkowski power of $\X$ is denoted $\X^n$. The identity matrix in $\R^{n\times n}$ is denoted $I_n$ (or $I$ when the dimension is clear), and the zero matrix in $\R^{m\times n}$ is denoted $0_{m\times n}$ (or $0$ when the dimension is clear).

\section{Background on Systems and Behaviors}
In this section, we formalize notions related to systems and specifications.

\begin{definition}\label{def:system}
A \emph{system} is a tuple $\S_\X=(\X,\U,f_\X)$ where $\X\subseteq\R^{n_\X}$ is a domain, $\U$ is an input set, and $f_\X:\R^{n_\X}\times\U\rarrows\R^{n_\X}$ is a set-valued dynamics of the form:
\begin{align}\label{eq:originalDynamics}
x(t+1) \in f_\X( x(t), u(t)).
\end{align}
A tuple $(x,u)\in\X^{[0;T[}\times\U^{[0;T[}$ is a \emph{solution} of the system $\S_\X$ (on $[0;T[$) if $T\in\mathbb{Z}_+\cup\{\infty\}$ and equation \eqref{eq:originalDynamics} holds for all $t\in[0;T-1[$.

A solution is \emph{maximal} when either (i) $T=\infty$, or (ii) $T<\infty$ and $f_\X(x(T-1),u(T-1))=\emptyset$, or (iii) $T<\infty$ and $f_\X(x(T-1),u(T-1))\nsubseteq\X$.
\end{definition}

Then, the behavior of a system is defined as the set of its maximal solutions.
\begin{definition} \label{def:sys:behavior}
The \emph{behavior} of a system $\S_\X$ is the set
\begin{align*}
\{(x,u)\mid \exists T: (x,u) \text{ is a solution of }\S_\X \text{ on }[0;T[\\ \text{ that is maximal.}\},
\end{align*}
and is denoted $\B[\S_\X]$. The \emph{closed-loop behavior} of a system $\S_\X$ under a policy $\pi:\X^*\rightarrow\U$ is the set 
$$
\{(x,u)\in \B[\S_\X] \mid u(t)=\pi(x(0),\dots,x(t))\ \forall t\in[0;T[ \},
$$
and is denoted $\B_\pi[\S_\X]$.
\end{definition}

Let us define the notion of specification.
\begin{definition}
A \emph{specification} over $\X\times\U$ is a set $S\subseteq(\X\times\U)^\infty$.
\end{definition}

Finally, we can formalize the notion of specification satisfaction.

\begin{definition}\label{def:sys:satisfaction}
The system $\S$ under the policy $\pi:\X^*\rightarrow\U$ \emph{satisfies} the specification $S$ if the closed-loop behavior of the system $\S$ under the policy $\pi$ is included in the specifications, i.e., $\B_\pi[\S]\subseteq S$. In that case, we write $\S\models_\pi S$.
\end{definition}

\newpage
\section{Lifted Systems and Simulations}\label{sec:methods}

When the system $\S$ is complex (e.g., nonlinear), finding a policy that satisfies a given specification is hard. To tackle this issue, we propose to abstract the system into a lifted system for which control design might be easier.

In this section, we first introduce the notion of lifted systems (Subsection~\ref{sec:liftedSystems}), and then a notion of simulation among them (Subsection~\ref{sec:simulation}).

\subsection{Lifted Systems}\label{sec:liftedSystems}

Lifted systems are defined as follows.

\begin{definition}\label{def:liftedSystem}
A \emph{lifted system} is a tuple $\LS_\Y =\\ (\X,\U,\psi_\Y,f_\Y,g_\Y)$ with:
\begin{itemize}
\item $\X\subseteq\R^{n_\X}$ an output set
\item $\U$ an input set
\item $\psi_\Y:\R^{n_\X}\rightarrow\R^{n_\Y}$ a lifting function
\item $f_\Y:\R^{n_\Y}\times\U\rarrows\R^{n_\Y}$ a dynamics
\item $g_\Y:\R^{n_\Y}\rightarrow\R^{n_\X}$ an output map such that for all $x\in\R^{n_\X}$, $g_\Y(\psi_\Y(x))=x$.
\end{itemize}
The \emph{dimension} of the lifted system is $n_\Y$. The lifted system is said to be \emph{affine} if $f_\Y(y,u)=A_\Y y+B_\Y u\oplus W_\Y$ and $g_\Y(y)=C_\Y y$ for some matrices $A_\Y$, $B_\Y$, $C_\Y$, and some (possibly unbounded) polyhedra $W_\Y$.
\end{definition}

We use the following notation: for a lifted system $\LS_\A$, its lifting function, dynamics, output map and dimension are denoted $\psi_\A$, $f_\A$, $g_\A$ and $n_\A$, respectively. Similarly, if the lifted system is affine, the quantities $A_\A$, $B_\A$, $C_\A$, and $W_\A$ describe its dynamics as in Definition~\ref{def:liftedSystem}. We note that since the output map $g_\Y$ is a left inverse of the lifting function $\psi_\Y$, it follows that $\psi_\Y$ is injective and $g_\Y$ is surjective.

An \emph{unlifted system} is a lifted system whose dimension is $n_\X$ and lifting function and output maps are the identity. We identify any unlifted system $\LS_\Y=(\X,\U,I,f_\Y,I)$ with the system $\S_\Y=(\X,\U,f_\Y)$ (see Definition~\ref{def:system}).

From a practical perspective, our goal is to abstract a nonlinear unlifted system (i.e., a system of the form \eqref{eq:originalDynamics}) into an affine lifted system. In addition, we want to be able to compare two affine lifted systems in terms of their ability to satisfy specifications.

\begin{definition}
A tuple $(x,u,y)\in\X^{[0;T[}\times\U^{[0;T[}\times(\R^{n_\Y})^{[0;T[}$ is a \emph{solution} of the lifted system $\LS_\Y$ (on $[0;T[$) if $T\in\mathbb{Z}_+\cup\{\infty\}$ and
\begin{align*}
y(0)&=\psi_\Y(x(0)) \\
y(t+1)&\in f_\Y(y(t),u(t))\ \forall t\in[0;T-1[ \\
x(t)&=g_\Y(y(t))\ \forall t\in [0;T[.
\end{align*}
A solution is \emph{maximal} when either (i) $T=\infty$, or (ii) $T<\infty$ and $f_\Y(y(T-1),u(T-1))=\emptyset$, or (iii) $T<\infty$ and $g_\Y(f_\Y(y(T-1),u(T-1)))\nsubseteq\X$.
\end{definition}

The state of the lifted system $\LS_\Y$ refers to $y(t)$ and its output refers to $x(t)$. The set of initial states is implicitly given by $y(0)\in\psi_\Y(\X)$.

Note that the condition $x(t)\in\X$ is not necessarily satisfied by $x(t)=g_\Y(y(t))$ since $y(t)\in\R^{n_\Y}$. For that reason, we define the \emph{domain} of the lifted system $\LS_\Y$ as the set of states $y$ that give an output in $\X$. It is written $\mathcal{D}_\Y\coloneqq g_\Y^{-1}[\X]=\{y\in\R^{n_\Y}\mid g_\Y(y)\in \X\}$.

The case (ii) for a solution to be maximal happens when the solution can not be extended. The case (iii) indicates that the state may leave the domain $\D_\Y$, i.e., $f_\Y(y(T-1),u(T-1))\nsubseteq\D_\Y$.

We define the behavior of a lifted system next.
\begin{definition}\label{def:behavior}
The \emph{behavior} of a lifted system $\LS_\Y$ is the set
\begin{align*}
\{(x,u)\mid \exists y,T: (x,u,y) \text{ is a solution of }\LS_\Y \text{ on }[0;T[\\ \text{ that is maximal.}\},
\end{align*}
and is denoted $\B[\LS_\Y]$. The \emph{closed-loop behavior} of a lifted system $\LS_\Y$ under a policy $\pi:\X^*\rightarrow\U$ is the set
$$
\{(x,u)\in \B[\LS_\Y] \mid u(t)=\pi(x(0),\dots,x(t))\ \forall t\in[0;T[ \},
$$
and is denoted $\B_\pi[\LS_\Y]$.
\end{definition}

Note that the feedback policy $\pi$ is a function of the previous outputs $x(t)$ and not a function of the previous states $y(t)$. This allows us to implement policies designed for the lifted system on the unlifted one, and compare closed-loop behaviors.

For a lifted system $\LS_\Y$ that is unlifted, the definition of behavior (Definition~\ref{def:behavior}) coincides with the one for systems (Definition~\ref{def:sys:behavior}).

Finally, the notion of specification satisfaction is defined similarly to one in Definition~\ref{def:sys:satisfaction}.
\begin{definition}
The lifted system $\LS$ under the policy $\pi:\X^*\rightarrow\U$ \emph{satisfies} the specification $S$ if the closed-loop behavior of the lifted system $\LS$ under the policy $\pi$ is included in the specification, i.e., $\B_\pi[\LS]\subseteq S$. In that case, we write $\LS\models_\pi S$.
\end{definition}

\subsection{Simulation between lifted systems}\label{sec:simulation}

In this section, we introduce the notion of simulation between lifted systems. Then, we prove that this relation implies the containment of open- and closed-loop behaviors of the lifted systems.

\begin{definition} \label{def:order}
Given two lifted systems $\LS_\Y$ and $\LS_\Z$ with same sets $\X$ and $\U$, we say that $\LS_\Z$ \emph{simulates} $\LS_\Y$ (or that $\LS_\Y$ \emph{refines} $\LS_\Z$), if there exists a set-valued function (called \emph{refinement map}) $\rho:\R^{n_\Z}\rarrows\R^{n_\Y}$ such that
\begin{subequations} \label{eq:order}
\begin{align}
\forall x\in\X:&\ \psi_\Y(x)\in\rho(\psi_\Z(x)) \label{eq:order:psi} \\
\forall (z,u)\in \D_\Z\times\U:&\ f_\Y(\rho(z),u) \subseteq \rho(f_\Z(z,u)) \label{eq:order:dyn} \\
\forall z\in \R^{n_\Z}:&\ g_\Y(\rho(z)) \subseteq \{g_\Z(z)\} \label{eq:order:output} \\
\forall (z,u)\in \D_\Z\times\U, \forall y\in&\rho(z):\ f_\Y(y,u)=\emptyset \Rightarrow f_\Z(z,u)=\emptyset. \label{eq:order:block}
\end{align}
\end{subequations}
In that case, we write $\LS_\Y\preceq\LS_\Z$ (or $\LS_\Y\preceq_\rho\LS_\Z$ to make the refinement map explicit).
\end{definition}

Since $g_\A$ is the left inverse of $\psi_\A$ for $\A\in\{\Y,\Z\}$, condition~\eqref{eq:order:psi} ensures that any pair of initial states $y(0)$ and $z(0)$ of $\LS_\Y$ and $\LS_\Z$ that give the same output, i.e., $g_\Y(y(0))=g_\Z(z(0))=x(0)$ are related through $\rho$, i.e., $y(0)\in\rho(z(0))$.

Condition \eqref{eq:order:dyn} ensures that if $y$ and $z$ are related through $\rho$, i.e., $y\in\rho(z)$, and if $z$ is in the domain $\D_\Z$, then, their successors $y^+$ and $z^+$ will be related in the same way, i.e., $y^+\in\rho(z^+)$. This condition can be rewritten using quantifiers as
\begin{align*}
\forall (z,u)\in \D_\Z\times\U, \forall y\in\rho(z), \forall y^+\in f_\Y(y,u),\\
\exists z^+\in f_\Z(z,u) \text{ s.t. } y^+\in \rho(z^+).
\end{align*}
This condition is similar to the one used for simulation relations \citep[Def.~4.7. Cond.~3.]{tabuada2009verification} or alternating simulation relations \citep[Eq.~(4)]{reissig2016feedback} with the exception that we additionally require the two inputs $u_\Y$ and $u_\Z$ to be the same, i.e., $u_\Y=u_\Z=u$.

Condition~\eqref{eq:order:output} states that if $y$ and $z$ are related through $\rho$, then they must have the same output, i.e., $g_\Y(y)=g_\Z(z)$. A similar condition can be found in \cite[Def.~3.1, (A3)]{majumdar2020abstraction}. Condition~\eqref{eq:order:output} implies that the domains of the two lifted systems satisfy
\begin{align}\label{eq:order:domain}
\rho(\D_\Z)\subseteq\D_\Y.
\end{align}
Indeed, if $z\in\D_\Z$, i.e., $g_\Z(z)\in\X$, then $g_\Y(\rho(z))\subseteq \{g_\Z(z)\}\subseteq \X$, which is $\rho(z)\subseteq\D_\Y$. Informally, \eqref{eq:order:domain} can be interpreted as the domain $\D_\Y$ being ``not smaller" than the domain $\D_\Z$.

Finally, \eqref{eq:order:block} states that if $z$ is in the domain and $y$ is related to $z$, then if $y$ is blocking for an input $u$, then $z$ must be blocking for the same $u$. A similar condition is considered in the context of feedback refinement for output feedback control \cite[Def.~V.2 (i)]{reissig2016feedback}. Note that if the dynamics $f_\Y(y,u)$ is non-blocking, then \eqref{eq:order:block} trivially holds.

\begin{rem}
One may think a good option for the refinement map $\rho$ may be $\rho(z) = \psi_\Y(\psi_\Z^{-1}[z])$. Indeed, it ensures that conditions \eqref{eq:order:psi} and \eqref{eq:order:output} necessarily hold. However, with this choice of refinement map, condition \eqref{eq:order:dyn} will generally not hold. To see why, note that for all $(z,u)$, the right-hand side of \eqref{eq:order:dyn} will be included in $\psi_\Y(\X)$. But for most $f_\Y$, the left-hand side of \eqref{eq:order:dyn} will not be included in $\psi_\Y(\X)$. In particular, it will not hold if $\LS_\Y$ is affine and $W_\Y$ has a non-zero volume.
\end{rem}

\begin{thm}
The relation $\preceq$ is a preorder, i.e., given three lifted systems $\LS_i$ for $i=1,2,3$, the following transitivity property holds: $\LS_1 \preceq_{\rho_{1\leftarrow 2}} \LS_2 \preceq_{\rho_{2\leftarrow 3}} \LS_3$ implies $\LS_1 \preceq_{\rho_{1\leftarrow 2}\circ\rho_{2\leftarrow 3}} \LS_3$; and the following reflexivity property holds: $\LS_i \preceq_I \LS_i$.
\end{thm}
\begin{pf}
The proof of the reflexivity is straightforward. To prove transitivity, let us denote $\rho_{1\leftarrow3}\coloneqq\rho_{1\leftarrow 2}\circ\rho_{2\leftarrow 3}$ and write $\eqref{eq:order}_{i\leftarrow j}$ to refer to the equations \eqref{eq:order} for $\LS_i\preceq_{\rho_{i\leftarrow j}}\LS_j$. Given $\eqref{eq:order}_{1\leftarrow 2}$ and $\eqref{eq:order}_{2\leftarrow 3}$, we need to prove $\eqref{eq:order}_{1\leftarrow 3}$.

For $\eqref{eq:order:psi}_{1\leftarrow 3}$: Let $x\in\X$. Using $\eqref{eq:order:psi}_{1\leftarrow 2}$ and $\eqref{eq:order:psi}_{2\leftarrow 3}$, we have $\psi_1(x)\in\rho_{1\leftarrow2}(\psi_2(x))\subseteq \rho_{1\leftarrow2}(\rho_{2\leftarrow3}(\psi_3(x)))$, which proves $\eqref{eq:order:psi}_{1\leftarrow 3}$.

For $\eqref{eq:order:dyn}_{1\leftarrow 3}$: Let $(z,u)\in\D_3\times\U$. Using $\eqref{eq:order:dyn}_{2\leftarrow 3}$, $f_2(\rho_{2\leftarrow3}(z),u)\subseteq\rho_{2\leftarrow3}(f_3(z,u))$ and then $\rho_{1\leftarrow2}(f_2(\rho_{2\leftarrow3}(z),\\ u))\subseteq\rho_{1\leftarrow2}(\rho_{2\leftarrow3}(f_3(z,u)))$. On the other hand, using \eqref{eq:order:domain}, $\rho_{2\leftarrow3}(z)\subseteq\D_2$, and $\eqref{eq:order:dyn}_{1\leftarrow 2}$ gives $f_1(\rho_{1\leftarrow2}(\rho_{2\leftarrow3}(z)),u)\subseteq \rho_{1\leftarrow2}(f_2(\rho_{2\leftarrow3}(z),u))$. We have shown that the right hand side is included in $\rho_{1\leftarrow3}(f_3(z,u))$, so $f_1(\rho_{1\leftarrow3}(z),u) \subseteq \rho_{1\leftarrow3}(f_3(z,u))$, which concludes the proof of $\eqref{eq:order:dyn}_{1\leftarrow3}$.

For $\eqref{eq:order:output}_{1\leftarrow 3}$: Let $z\in\R^{n_3}$. From $\eqref{eq:order:output}_{1\leftarrow 2}$, $g_1(\rho_{1\leftarrow2}(\rho_{2\leftarrow3}(z)))\\ \subseteq g_2(\rho_{2\leftarrow3}(z))$ and from $\eqref{eq:order:output}_{2\leftarrow 3}$, the right hand side is included in $\{g_3(z)\}$, which shows $\eqref{eq:order:output}_{1\leftarrow 3}$ holds.

For $\eqref{eq:order:block}_{1\leftarrow 3}$: Let $(z_3,u)\in\D_3\times\U$ and let $z_1\in\rho_{1\leftarrow2}(\rho_{2\leftarrow3}(z_3))$. Then, there exists $z_2\in\rho_{2\leftarrow3}(z_3)$ such that $z_1\in\rho_{1\leftarrow2}(z_2)$. It follows from \eqref{eq:order:domain} that $z_2\in\D_2$. Then, $\eqref{eq:order:block}_{1\leftarrow 2}$ and $\eqref{eq:order:block}_{2\leftarrow 3}$ imply that
if $f_1(z_1,u)=\emptyset$, then $f_2(z_2,u)=\emptyset$, and then $f_3(z_3,u)=\emptyset$; proving $\eqref{eq:order:block}_{1\leftarrow 3}$.\hspace*{\fill}\qed
\end{pf}

The next theorem states that the open-loop behavior of a lifted system is included in the behavior of its simulation.

\begin{thm}\label{thm:behavior}
Given two lifted systems $\LS_\Y$ and $\LS_\Z$, if $\LS_\Y$ refines $\LS_\Z$, then the behavior of $\LS_\Y$ is included in the behavior of $\LS_\Z$, i.e., $\LS_\Y\preceq \LS_\Z$ implies $\B[\LS_\Y]\subseteq \B[\LS_\Z]$.
\end{thm}

\begin{pf}
Assume that $\LS_\Y\preceq_\rho\LS_\Z$ and let $(x,u)\in\B[\LS_\Y]$. Then, there exists $y$ and $T$ such that $(x,u,y)$ is a maximal solution of $\LS_\Y$ on $[0;T[$. We need to show that there exists $z$ such that $(x,u,z)$ is a maximal solution of $\LS_\Z$ on $[0;T[$.

First, let us prove the following statement (\textbf{S}): for all $t\in[0;T-1[$, if there exists a $z\in\R^{n_\Z}$ such that $y(t)\in\rho(z)$ and $g_\Z(z)=x(t)$, then there exists $z^+\in f_\Z(z,u(t))$ such that $y(t+1)\in\rho(z^+)$ and $g_\Z(z^+)=x(t+1)$. To prove (\textbf{S}), note that $g_\Z(z)=x(t)\in\X$ implies $z\in \D_\Z$. Then, using \eqref{eq:order:dyn}, one can write
$$
y(t+1)\in f_\Y(y(t),u(t))\subseteq f_\Y(\rho(z),u(t))\subseteq \rho(f_\Z(z,u(t))),
$$
which implies that there exists $z^+\in f_\Z(z,u(t))$ such that $y(t+1)\in\rho(z^+)$. Then, using \eqref{eq:order:output}, one can write
$$
x(t+1)=g_\Y(y(t+1))\in g_\Y(\rho(z^+)) \subseteq \{g_\Z(z^+)\},
$$
and $g_\Z(z^+)=x(t+1)$. This concludes the proof of (\textbf{S}).

Because $z(0)\coloneqq\psi_\Z(x(0))$ satisfies $y(0)\in\rho(z(0))$ (thanks to \eqref{eq:order:psi}) and $g_\Z(z(0))=x(0)$ (because $g_\Z$ is a left inverse of $\psi_\Z$), it follows recursively from (\textbf{S}) that there exists a sequence $z\in(\R^{n_\Z})^{[0;T[}$ such that $(x,u,z)$ is a solution of $\LS_\Z$ on $[0;T[$ and $y(t)\in\rho(z(t))$ for all $t\in[0;T[$.

It remains to prove that the solution $(x,u,z)$ is maximal. Since $(x,u,y)$ is maximal, either (i) $T=\infty$, and $(x,u,z)$ is maximal; or (ii) $T<\infty$ and $f_\Y(y(T-1),u(T-1))=\emptyset$. Since $y(T-1)\in\rho(z(T-1))$, \eqref{eq:order:block} implies $f_\Z(z(T-1),u(T-1))=\emptyset$ and $(x,u,z)$ is maximal; or (iii) $T<\infty$ and $g_\Y(f_\Y(y(T-1),u(T-1)))\nsubseteq\X$. It follows from $y(T-1)\in\rho(z(T-1))$ and \eqref{eq:order:dyn} that
\begin{align*}
f_\Y(y(T-1),u(T-1))&\subseteq f_\Y(\rho(z(T-1)),u(T-1)) \\
&\subseteq \rho( f_\Z(z(T-1),u(T-1) )).
\end{align*}
Then, using \eqref{eq:order:output},
\begin{align*}
g_\Y(f_\Y(y(T-1),u(T&-1))) \\
&\subseteq g_\Y( \rho( f_\Z(z(T-1),u(T-1)) )) \\
&\subseteq g_\Z( f_\Z(z(T-1),u(T-1)) ),
\end{align*}
which implies $g_\Z( f_\Z(z(T-1),u(T-1)) )\nsubseteq\X$, and $(x,u,z)$ is maximal. This concludes the proof. \hspace*{\fill}\qed
\end{pf}

A consequence of Theorem~\ref{thm:behavior} is the inclusion of closed-loop behaviors.

\begin{thm}\label{thm:closedLoopBahavior}
Given two lifted systems $\LS_\Y$, $\LS_\Z$, and a policy $\pi:\X^*\rightarrow\U$, if $\LS_\Y$ refines $\LS_\Z$, then the closed-loop behavior of $\LS_\Y$ under $\pi$ is included in the closed-loop behavior of $\LS_\Z$ under $\pi$, i.e., $\LS_\Y\preceq\LS_\Z$ implies $\B_\pi[\LS_\Y]\subseteq \B_\pi[\LS_\Z]$.
\end{thm}
\begin{pf}
For a given policy $\pi:\X^*\rightarrow\U$, define the set 
\begin{align*}
\Pi\coloneqq\bigcup_{T\in\mathbb{Z}_+\cup \{\infty\}} \{& (x,u)\in\X^{[0;T[}\times\U^{[0;T[}\mid \\
&u(t)=\pi(x(0),\dots,x(t)),\ \forall t\in[0;T[ \}.
\end{align*}
By definition of the closed loop behavior, $\B_\pi[\LS_\A]=\B[\LS_\A]\cap\Pi$ for $\A\in\{\Y,\Z\}$. By Theorem~\ref{thm:behavior}, $\LS_\Y\preceq \LS_\Z$ implies $\B[\LS_\Y]\subseteq \B[\LS_\Z]$. Then, $\B[\LS_\Y]\cap\Pi\subseteq \B[\LS_\Z]\cap\Pi$, concluding the proof.\hspace*{\fill}\qed
\end{pf}

A direct consequence of Theorem~\ref{thm:closedLoopBahavior} is that if an abstract lifted system under a policy satisfies some specification, then the concrete system under the same policy satisfies this specification as well.

\begin{cor}
Given two lifted systems $\LS_\Y$, $\LS_\Z$, a policy $\pi:\X^*\rightarrow\U$, and a specification $S$ over $\X\times\U$, if $\LS_\Y$ refines $\LS_\Z$, and if $\LS_\Z$ satisfies the specification $S$ under the policy $\pi$, then $\LS_\Y$ satisfies the same specification $S$ under the same policy $\pi$, i.e., $\LS_\Y\preceq\LS_\Z\models_\pi S$ implies $\LS_\Y\models_\pi S$.
\end{cor}

Informally, if $\LS_\Y$ and $\LS_\Z$ are two abstractions of the same concrete system of interest $\LS_\X$, and if $\LS_\Y\preceq\LS_\Z$, then $\LS_\Y$ is a ``not worse'' representation of $\LS_\X$ than $\LS_\Z$ in terms of specification satisfaction.

\subsection{Simulating unlifted systems}
In this Section, we focus on the special case where the simulated system is unlifted, i.e., the simulated system can be identified to a system according to Definition~\ref{def:system}. This case is of particular interest since in practice, we are interested in controlling systems of the form \eqref{eq:originalDynamics}. The next theorem gives a sufficient condition to simulate an unlifted system.

\begin{thm} \label{thm:unlifted}
Given an unlifted system $\LS_\X$ and a lifted system $\LS_\Z$, $\LS_\Z$ simulates $\LS_\X$ with refinement map $\psi_\Z^{-1}$, i.e., $\LS_\X\preceq_{\psi^{-1}_\Z}\LS_\Z$, if and only if $\forall(x,u)\in \X\times\U$:
\begin{subequations} \label{eq:unlifted}
\begin{align}
\psi_\Z(f_\X(x,u)) &\subseteq f_\Z(\psi_\Z(x),u), \label{eq:unlifted:dyn} \\
f_\X(x,u)=\emptyset &\Rightarrow f_\Z(\psi_\Z(x),u)=\emptyset. \label{eq:unlifted:block}
\end{align}
\end{subequations}
\end{thm}
\begin{pf}
First, \eqref{eq:order:psi} reduces to $x\in\psi_\Z^{-1}[\psi_\Z(x)]$ which holds trivially. Similarly, \eqref{eq:order:output} reduces to $\psi_\Z^{-1}[z]\subseteq\{g_\Z(z)\}$ which holds since $g_\Z$ is a left inverse of $\psi_\Z$. Indeed, if $x\in\psi_\Z^{-1}[z]$, then $z=\psi_\Z(x)$ and $g_\Z(z)=g_\Z(\psi_\Z(x))=x$.

Conditions \eqref{eq:order:dyn} and \eqref{eq:order:block} reduce to, $\forall (z,u)\in\D_\Z\times\U:$
\begin{subequations} \label{eq:unlifted:aux}
\begin{align}
\ f_\X(\psi_\Z^{-1}[z],u)&\subseteq\psi_\Z^{-1}[f_\Z(z,u)] \label{eq:unlifted:aux:dyn} \\
\forall x\in\psi_\Z^{-1}[z]:\ f_\X(x,u)=\emptyset &\Rightarrow f_\Z(z,u)=\emptyset. \label{eq:unlifted:aux:block}
\end{align}
\end{subequations}

It remains to prove \eqref{eq:unlifted} if and only if \eqref{eq:unlifted:aux}. Let us prove that \eqref{eq:unlifted} implies \eqref{eq:unlifted:aux}. To this end, let $(z,u)\in\D_\Z\times\U$. If $\psi_\Z^{-1}[z]=\emptyset$, then \eqref{eq:unlifted:aux} vacuously hold. Otherwise, let $x\in\R^{n_\X}$ be such that $\psi_\Z(x)=z$. Since $z\in\D_\Z$, $\X\ni g_\Z(z)=g_\Z(\psi_\Z(x))=x$ and \eqref{eq:unlifted:block} implies \eqref{eq:unlifted:aux:block}. In addition, \eqref{eq:unlifted:dyn} gives $\psi_\Z(f_\X(\psi_\Z^{-1}[z],u))\subseteq f_\Z(z,u)$ and applying $\psi_\Z^{-1}$ on both sides gives \eqref{eq:unlifted:aux:dyn}.

Finally, let us prove that \eqref{eq:unlifted:aux} implies \eqref{eq:unlifted}. Let $(x,u)\in\X\times\U$ and define $z\coloneqq\psi_\Z(x)\in\D_\Z$. Then, \eqref{eq:unlifted:aux:block} implies \eqref{eq:unlifted:block}. In addition, \eqref{eq:unlifted:aux:dyn} gives $f_\X(x,u)\subseteq\psi_\Z^{-1}[f_\Z(\psi_\Z(x),u)]$. By applying $\psi_\Z$ on both sides, we have
$$
\psi_\Z(f_\X(x,u))\subseteq\psi_\Z(\psi_\Z^{-1}[f_\Z(\psi_\Z(x),u)])\subseteq f_\Z(\psi_\Z(x),u),
$$
which is \eqref{eq:unlifted:dyn}, concluding the proof.\hspace*{\fill}\qed
\end{pf}

As we will see in Section~\ref{sec:computation:unlifted}, Theorem~\ref{thm:unlifted} is of practical interest when one wants to simulate an unlifted system since conditions \eqref{eq:unlifted:dyn} can be relaxed into sum-of-square constraints. However, note that even if $\LS_\X\npreceq_{\psi_\Z^{-1}}\LS_\Z$, there may exist another refinement map ensuring $\LS_\X\preceq\LS_\Z$.

If $\LS_\Z$ is affine, condition \eqref{eq:unlifted:dyn} is the definition of Koopman over-approximation \cite[Def.~2]{balim2023koopman}. If in addition, no control is considered, i.e., $\U=\{0\}$, then it is the definition of approximate immersion in \citep[Def.~5]{wang2023computation}.

\section{Verifying $\preceq$}\label{sec:verification}

In this section, we derive optimization-based sufficient conditions to guarantee that one lifted system simulates another. First, we consider the case of an affine lifted system with polynomial lifting simulating a deterministic unlifted polynomial system. Then, we consider the case of two affine lifted systems with polynomial lifting functions.

Given an unlifted system, this allows to compute simulating affine lifted systems with different lifting functions, and then check if one lifted system refines the other.

\subsection{Finding a simulating affine lifted system} \label{sec:computation:unlifted}
We show how to find an affine lifted system $\LS_\Z$ simulating a deterministic unlifted system $\LS_\X$. We assume that the lifting function $\psi_\Z$ is given, and our goal is to find a dynamics $f_\Z$ such that $\LS_\X\preceq\LS_\Z$. In this subsection, we make the following assumption to make this problem tractable:
\begin{assumption} \label{assumption:computation:unlifted}
$\ $

\begin{itemize}
\item The set $\X\times\U$ can be written $\{(x,u)\in\R^{n_\X}\times\R^{n_\U}\mid l_{j}(x,u)\geq0, j\in \mathcal{J}\}$ for some polynomials $l_{j}:\R^{n_\X\times n_\U}\rightarrow \R$ and some finite set of indices $\mathcal{J}$.
\item The unlifted system $\LS_\X$ is such that
\begin{itemize}
\item The dynamics $f_\X$ is single-valued, i.e., it is identified with a function $f_\X:\R^{n_\X}\times\U\rightarrow\R^{n_\X}$.
\item The dynamics $f_\X$ is polynomial.
\end{itemize}
\item The lifted system $\LS_\Z$ is such that
\begin{itemize}
\item The dynamics $f_\Z$ is affine with non-empty $W_\Z$.
\item The lifting function $\psi_\Z$ is polynomial.
\end{itemize}
\end{itemize}
\end{assumption}

According to Theorem \ref{thm:unlifted}, a sufficient condition for $\mathsf{LifSys}_\X\preceq\mathsf{LifSys}_\Z$ is \eqref{eq:unlifted}. Since, $f_\X$ is single valued, $f_\X(x,u)\neq\emptyset$ for all $(x,u)$ and \eqref{eq:unlifted:block} holds. Using the H-representation of the polytope $W_\Z$, i.e., $W_\Z=\{z\mid H_{\Z,i} z\leq h_{\Z,i}, \text{ for all } i\in \mathcal{I}\}$, condition \eqref{eq:unlifted:dyn} can be rewritten: for all $(x,u)\in \X\times\U$ and for all $i\in \mathcal{I}$,
$$
h_{\Z,i}- H_{\Z,i}\big(\psi_\Z(f_\X(x,u))-A_\Z\psi_\Z(x)-B_\Z u \big) \geq0.
$$
A sufficient condition is the existence of polynomials $\lambda_{i,j}(x,u)$ such that (see e.g., \cite[Eq.~(35)]{papachristodoulou2005tutorial})
\begin{subequations}\label{eq:unlifted:sos}
\begin{align}
&\forall (i,j)\in \mathcal{I}\times \mathcal{J}:\ \lambda_{i,j}(x,u) \geq 0 \\
&\forall i\in \mathcal{I}:\ h_{\Z,i}- H_{\Z,i}\big(\psi_\Z(f_\X(x,u))-A_\Z\psi_\Z(x)-B_\Z u \big) \notag \\
&\hspace{2cm}- \sum_{j\in \mathcal{J}} \lambda_{i,j}(x,u)l_{j}(x,u) \geq0.
\end{align}
\end{subequations}
Under Assumption~\ref{assumption:computation:unlifted}, conditions~\eqref{eq:unlifted:sos} are polynomial positivity constraints that can be relaxed into sum-of-squares (SOS) constraints. For given $A_\Z$, $B_\Z$ and $W_\Z$, this relaxation results in a semidefinite program in $\lambda_{i,j}$ whose feasibility is then a sufficient condition for $\mathsf{LifSys}_\X\preceq\mathsf{LifSys}_\Z$.

Furthermore, $A_\Z$, $B_\Z$, $h_\Z$ and $\lambda_{i,j}$ can be co-optimized (note that the $H_W$ matrix can not be optimized without causing the SOS program to be non-convex). To make the nondeterminism as small as possible in $f_\Z$, we solve
\begin{align}\label{eq:unlifted:optiProb}
\min_{A_\Z,\ B_\Z,\ h_\Z,\ \lambda_{i,j}} \sum_{i\in \mathcal{I}} | h_{\Z,i} | \text{ s.t. } \eqref{eq:unlifted:sos}.
\end{align}
A solution to this convex problem gives an affine dynamics $f_\Z$ such that $\mathsf{LifSys}_\X\preceq\mathsf{LifSys}_\Z$.

\subsection{Comparing two affine lifted systems} \label{sec:computation:affine}

Given two affine lifted systems $\mathsf{LifSys}_\Y$ and $\mathsf{LifSys}_\Z$, we want to verify if $\mathsf{LifSys}_\Y\preceq\mathsf{LifSys}_\Z$. Note that finding a $\rho$ satisfying $\eqref{eq:order}$ is an infinite dimensional feasibility problem for the following reasons: (i) the set of set-valued refinement maps $\rho:\R^{n_\Z}\rarrows\R^{n_\Y}$ is infinite dimensional; and (ii) each of the four conditions in \eqref{eq:order} contain a for all quantifier ($\forall$), resulting in an infinite number of constraints. To make this problem tractable, we make the following assumption:
\begin{assumption} \label{assumption:computation:affine}
$\ $

\begin{itemize}
\item The sets $\X$ and $\U$ are full-dimensional polytopes.
\item For $\A\in\{\Y,\Z\}$, the lifted system $\LS_\A$ is such that
\begin{itemize}
\item The lifting function $\psi_\A$ can be written $\psi_\A(x)=\begin{bmatrix}x \\ \psi_\A^{\bar{x}}(x)\end{bmatrix}$, with $\psi_\A^{\bar{x}}:\R^{n_\X}\rightarrow\R^{n_\A-n_\X}$.
\item The output map is $g_\A(z)=C_\A z$, with $C_\A=\begin{bmatrix} I_{n_\X} & 0_{n_\X\times (n_\A-n_\X)}\end{bmatrix}$.
\item The lifting function $\psi_\A$ is polynomial.
\item The dynamics $f_\A$ is affine with full-dimensional $W_\A$.
\end{itemize}
\end{itemize}
\end{assumption}
Finally, we restrict our attention to \emph{affine refinement maps} $\rho$, i.e., maps that can be written
$$
\rho(z)=Rz\oplus W_\rho,
$$
with $R\in\R^{n_\Y\times n_\Z}$ and $W_\rho\subseteq\R^{n_\Y}$ a non-empty and bounded polytope.

This allows to rewrite the refinement conditions as a finite-dimensional feasibility problem. To this end, let us introduce the following notation: for a matrix $A\in\R^{m\times n}$, its block decompositions are written
$$
A=\begin{bmatrix} A^{xx} & A^{x\bar{x}} \\ A^{\bar{x}x} & A^{\bar{x}\bar{x}}\end{bmatrix}
=\begin{bmatrix} A^{:x} & A^{:\bar{x}} \end{bmatrix}
=\begin{bmatrix} A^{x:} \\ A^{\bar{x}:} \end{bmatrix},
$$
where $A^{xx}\in\R^{n_\X\times n_\X}$, $A^{:x}\in\R^{m\times n_\X}$ and $A^{x:}\in\R^{n_\X\times n}$.

\begin{thm} \label{thm:order:affine}
Given two affine lifted systems $\mathsf{LifSys}_\Y$ and $\mathsf{LifSys}_\Z$ satisfying Assumption~\ref{assumption:computation:affine}. There exists an affine refinement map $\rho$ such that $\mathsf{LifSys}_\Y\preceq_\rho\mathsf{LifSys}_\Z$ if and only if there exists a matrix $R\in\R^{n_\Y\times n_\Z}$ and a non-empty bounded polytope $W_\rho^{\bar{x}}\subseteq\R^{n_\Y-n_\X}$ such that
\begin{subequations}\label{eq:order:affine}
\begin{align}
R^{x:}&=\begin{bmatrix} I & 0 \end{bmatrix} \label{eq:order:affine:simple} \\
\forall x\in \X:\ \psi^{\bar{x}}_\Y(x) -R^{\bar{x}:}\psi_\Z(x) &\in W^{\bar{x}}_\rho \label{eq:order:affine:psi}\\
(A_\Y R-RA_\Z)^{:x}\X \oplus (B_\Y-RB_\Z)\U \oplus A_\Y^{:\bar{x}}& W^{\bar{x}}_\rho \notag\\
 \oplus W_\Y \subseteq RW_\Z \oplus (\{0_{n_\X}\}&\times W_\rho^{\bar{x}}) \label{eq:order:affine:dyn:x} \\
(A_\Y R-RA_\Z)^{:\bar{x}} &= 0. \label{eq:order:affine:dyn:x_bar}
\end{align}
\end{subequations}
In that case, $\rho(z)=\{z_{1:n_\X}\}\times ( R^{\bar{x}:}z\oplus W_\rho^{\bar{x}})$ is a refinement map.
\end{thm}
\begin{pf}
First, let us prove that an affine refinement map $\rho(z)=Rz\oplus W_\rho$ satisfies \eqref{eq:order:output} if and only if it can be written $\rho(z)=\{z_{1:n_\X}\}\times ( R^{\bar{x}:} z\oplus W^{\bar{x}}_\rho)$, i.e., equation \eqref{eq:order:affine:simple} holds and $W_\rho=\{0_{n_\X}\}\times W^{\bar{x}}_\rho$ for some polytope $W^{\bar{x}}_\rho$. To this end, let $z=\begin{bmatrix}x^\top & \bar{x}^\top\end{bmatrix}^\top\in\R^{n_\Z}$, with $x\in\R^{n_\X}$. Under Assumption~\ref{assumption:computation:affine}, condition~\eqref{eq:order:output} is
\begin{align*}
C_\Y(Rz\oplus W_\rho)&\subseteq \{C_\Z z\} \\
\Leftrightarrow R^{xx}x+R^{x\bar{x}}\bar{x}\oplus C_\Y W_\rho & \subseteq\{x\}.
\end{align*}
Since $W_\rho$ is non empty, $C_\Y W_\rho$ must be a singleton $\{r\}\subseteq\R^{n_\X}$. Then, $R^{xx}x+R^{x\bar{x}}\bar{x}+ r=x$ must hold for all $x$ and $\bar{x}$. This leads to $R^{xx}=I$, $R^{x\bar{x}}=0$ (which is \eqref{eq:order:affine:simple}), and $r=0$. Finally, $C_\Y W_\rho=\{0\}$ implies $W_\rho=\{0_{n_\X}\}\times W_\rho^{\bar{x}}$.

Then, using Assumption~\ref{assumption:computation:affine}, \eqref{eq:order:psi} is equivalent to \eqref{eq:order:affine:psi}.

Since $W_\Y\neq\emptyset$ (from Assumption~\ref{assumption:computation:affine}), for all $(y,u)\in\R^{n_\Y}\times\U$, $f_\Y(y,u)\neq\emptyset$ and \eqref{eq:order:block} vacuously holds.

Finally, let us prove that \eqref{eq:order:dyn} is equivalent to \eqref{eq:order:affine:dyn:x} and \eqref{eq:order:affine:dyn:x_bar}. Condition~\eqref{eq:order:dyn} can be rewritten: for all $(z,u)\in\D_\Z\times\U$,
\begin{align*}
A_\Y(Rz\oplus W_\rho)\oplus B_\Y u\oplus W_\Y \subseteq R(A_\Z z +B_\Z u\oplus W_\Z)\oplus W_\rho \\
\Leftrightarrow (A_\Y R-RA_\Z)z \oplus (B_\Y-RB_\Z)u \oplus A_\Y W_\rho \oplus W_\Y \\
\subseteq RW_\Z \oplus W_\rho,
\end{align*}
which holds if and only if
\begin{align}
(A_\Y R-RA_\Z)\D_\Z \oplus (B_\Y-RB_\Z)\U \oplus A_\Y^{:\bar{x}} W^{\bar{x}}_\rho \oplus W_\Y \notag \\
\subseteq RW_\Z \oplus W_\rho. \label{eq:affine:inclusion:proof}
\end{align}
Then, note that $g_\Z(z)=C_\Z z$ implies $\D_\Z=\X\times\R^{n_\Z-n_\X}$. Consequently, \eqref{eq:affine:inclusion:proof} is equivalent to \eqref{eq:order:affine:dyn:x} and \eqref{eq:order:affine:dyn:x_bar} (using the boundedness of $W_{\rho}$). This concludes the proof.\hspace*{\fill}\qed
\end{pf}

The practical importance of Theorem~\ref{thm:order:affine} lies in the following observations: (i) equations~\eqref{eq:order:affine:simple} and \eqref{eq:order:affine:dyn:x_bar} are linear in $R$, (ii) condition \eqref{eq:order:affine:psi} is a polynomial positivity constraint that can be handled using SOS optimization (in a similar way as in Section~\ref{sec:computation:unlifted}), and (iii) condition \eqref{eq:order:affine:dyn:x} is a polytope containment constraint, which can be handled via vertex enumeration (see Appendix~\ref{appendix:vertexEnumeration}), leading to bilinear constraints. Overall, these lead to a non-convex --- but finite-dimensional --- feasibility problem whose feasibility guarantees that $\LS_\Z$ simulates $\LS_\Y$, i.e., $\LS_\Y\preceq\LS_\Z$.

\begin{rem}
The methods developed in Section~\ref{sec:verification} can be adapted straightforwardly to \emph{picewise affine} lifted systems. That is lifted systems with dynamics $f(y,u)=A_{\sigma(C y)}y + B_{\sigma(C y)}u\oplus W_{\sigma(C y)}$, where $\sigma:\X\rightarrow\{1,\dots,K\}$ induces a partition $\{\X_k\}_{k=1}^{K}$ of $\X$. In the special case of a picewise affine unlifted system simulating a deterministic nonlinear unlifted system, condition~\eqref{eq:unlifted:dyn} reduces to: for $k=1,\dots,K$,
$$
\forall (x,u)\in\X_k\times\U:\ f_\X(x,u)\in A_k x + B_k u\oplus W_k,
$$
which is the definition of hybridization in \cite[Def.~3.2]{girard2011synthesis}.
\end{rem}

\section{Experiments and Discussion}
To demonstrate our method,\footnote{A \textsc{Julia} code that implements our method and generates the figure is available at \url{https://github.com/aaspeel/simLifSys}.} we consider the unlifted system $\LS_\X$ with dynamics $f_\X$ given by the Duffing equation $\ddot{x}=2x-2x^3-0.5\dot{x}+u$ discretized using explicit Euler with time step $0.1$. The state is $\begin{bmatrix}x & \dot{x}\end{bmatrix}^\top$, the set $\X$ is such that $-2\leq x\leq 2$ and $-3\leq \dot{x}\leq 3$, and the input set is $\U=[-50,50]$. Note that $\LS_\X$ satisfies Assumption~\ref{assumption:computation:unlifted}.

We consider the lifting functions $\psi_1=\begin{bmatrix}x & \dot{x}\end{bmatrix}^\top$, $\psi_2=\begin{bmatrix}x & \dot{x} & x^2 \end{bmatrix}^\top$ and $\psi_3=\begin{bmatrix}x & \dot{x} & x^2 &  x^3 \end{bmatrix}^\top$ and compute affine dynamics $f_1$, $f_2$ and $f_3$ by solving \eqref{eq:unlifted:optiProb} (with $H_{W_i}$ being the basis of an axis aligned hyper-box). For each $i\in\{1,2,3\}$, it gives an affine lifted system $\LS_i$ satisfying Assumption~\ref{assumption:computation:affine} such that $\LS_\X\preceq_{\psi_i^{-1}}\LS_i$. Then, we follow the method described in Section~\ref{sec:computation:affine} to verify if $\LS_i\preceq\LS_j$ for all $i\neq j\in\{1,2,3\}$. Unfortunately, the method was inconclusive. This can be because $\LS_i\npreceq\LS_j$ or it can be due to the conservativeness of our numerical method. Then, we inflate the sets $W_i$ of each lifted system $\LS_i$ by adding 10 to each component of the vector $h_i$. This leads to three other lifted systems $\widetilde{\LS}_i$ for $i\in\{1,2,3\}$. In that case, our method could verify that $\LS_i\preceq\widetilde{\LS}_j$ for all $i\in\{2,3\}$ and $j\leq i$.

\section{Conclusion and Future Directions}

This work introduces a new notion of simulation between lifted control systems. The proposed simulation relation is defined for general set-valued nonlinear and non-autonomous dynamics and is shown to imply containment of both open-loop and closed-loop behaviors. This new relation unifies and generalizes Koopman over-approximations~\citep{balim2023koopman}, approximate immersions~\citep{wang2023computation}, and hybridizations~\citep{girard2011synthesis}. Finally, optimization-based sufficient conditions are derived to verify if one lifted system simulates another. These contributions provide a theoretical and algorithmic framework for analyzing lifted systems, and finite-dimensional Koopman approximations in particular. In the future, we will derive less conservative conditions to verify the simulation relation. In addition, we will extend our approach to continuous-time and hybrid systems, non-polynomial but Lipschitz systems, and develop algorithms to compute ``good" lifting functions.

\emph{Acknowledgments:} The authors thank Zexiang Liu for insightful discussions on earlier versions of this work.

\bibliography{references_short} 

\appendix
\section{Vertex enumeration for (\ref{eq:order:affine:dyn:x})} \label{appendix:vertexEnumeration}
To handle condition $\eqref{eq:order:affine:dyn:x}$, we require that each vertex of the left-hand side polytope
$$
(A_\Y R-RA_\Z)^{:x}\X \oplus (B_\Y-RB_\Z)\U \oplus A_\Y W_\rho  \oplus W_\Y 
$$
be included in the right-hand side polytope
$$
RW_\Z \oplus W_\rho,
$$
where we write $W_\rho\coloneqq\{0_{n_\X}\}\times W_\rho^{\bar{x}}$.

Then, condition~\eqref{eq:order:affine:dyn:x} can be written
\begin{align*}
(A_\Y R-RA_\Z)^{:x}v^i_\X + (B_\Y-RB_\Z)v^j_\U +& A_\Y^{:\bar{x}} v^k_\rho \\
+ v^l_\Y &= Rw_\Z^{i,j,k,l} + w_\rho^{i,j,k,l} \\
H_{W_\Z} w_\Z^{i,j,k,l} &\leq h_{W_\Z} \\
H_{W_\rho} w_\rho^{i,j,k,l} &\leq h_{W_\rho}
\end{align*}
for each $v^i_\X$, $v^j_\U$, $v^k_\rho$, $v^l_\Y$ in the sets of vertices of $\X$, $\U$, $W_\rho$, and $W_\Y$, respectively. Note the bilinearities between $R$ and $w_\Z^{i,j,k,l}$.

Assuming that $W_\rho$ is an axis-aligned box, i.e., $W_\rho=\{y\mid h_{\underline{i}}\leq y_i\leq h_{\bar{i}} \}=\bigtimes_{i=1}^{n_\Y}[ h_{\underline{i}} , h_{\bar{i}} ]$, the vertices of $W_\rho$ are given by the vectors $v_{\rho}=\begin{bmatrix} h_{1^*}&\dots & h_{n_\Y^*} \end{bmatrix}^\top$ for each combination of $i^*\in\{\underline{i},\bar{i}\}$. Importantly, these vertices depend linearly on $h=h_{W_\rho}$.

Overall, this allows to write condition~\eqref{eq:order:affine:dyn:x} as bilinear constraints when $W_\rho$ is an axis align box.

\end{document}